\documentstyle[12pt,twoside,fleqn,espcrc1]{article}

\begin{document}
\setlength{\baselineskip}{3.0ex}
\setlength{\topmargin}{-15mm}

\title{Radiative and non radiative muon capture on the proton in heavy baryon
chiral perturbation theory
\thanks{This work was supported in part by the Deutsche Forschungsgemeinschaft,
the Natural Sciences and Engineering Research Council of Canada, and the 
NATO International Scientific Exchange Program. }}

\author{Harold W. Fearing $^a$, Randy Lewis $^a$, 
Nader Mobed $^b$, and Stefan Scherer $^c$}

\maketitle

\noindent $^a$ {TRIUMF, 4004 Wesbrook Mall, Vancouver, British Columbia, 
          Canada  V6T 2A3}\\[-1ex]

\noindent $^b$ {Department of Physics, University of Regina, Regina, 
          Saskatchewan, Canada S4S 0A2}\\[-1ex]

\noindent $^c$ {Institut f\"ur Kernphysik, Johannes Gutenberg-Universit\"at,
         J. J. Becher-Weg 45, D-55099 Mainz, Germany}\\[-1ex]

\begin{abstract}
We have evaluated the amplitude for muon capture by a proton, $\mu + p
\rightarrow n + \nu$, to ${\cal O}(p^3)$ within the context of heavy baryon
chiral perturbation theory (HBChPT) using the new ${\cal O}(p^3)$ Lagrangian of
Ecker and Moj\v{z}i\v{s} (E\&M). We obtain expressions for the standard muon
capture form factors and determine three of the coefficients of the E\&M
Lagrangian, namely, $b_7, b_{19}$, and $b_{23}$.  We describe progress on the
next step, a calculation of the radiative muon capture process, $\mu + p
\rightarrow n + \nu + \gamma$.
\end{abstract} \\

\section{INTRODUCTION}
Chiral perturbation theory is an effective theory for QCD, formulated in terms
of a series of effective Lagrangians of increasing orders in the momentum and
quark mass expansion.  It was originally formulated for mesons only, but can be
extended as HBChPT to include heavy baryons. For a review see
e.g. \cite{bkm,Ecker95,Pich}.

The complete Lagrangian for a single nucleon coupling to pions and external
fields up to third order in small momenta (denoted ${\cal L}^{EckM}_{\pi N}$)
has only recently been constructed by Ecker and Moj\v{z}i\v{s} \cite{EM}
(E\&M), although calculations with earlier versions and for specific processes
had been performed before.  Here we study muon capture by a proton with the new
Lagrangian, ${\cal L}^{EckM}_{\pi N}$.  The form factors that appear in the
muon capture amplitude have been considered previously within HBChPT, but not
with the new Lagrangian ${\cal L}^{EckM}_{\pi N}$.

Our calculation gives explicit expressions for each of the muon capture form
factors, in terms of parameters that appear in ${\cal L}^{EckM}_{\pi N}$.  We
use experimental data to determine the numerical values of the parameters,
which are directly transferable to future calculations of other processes where
${\cal L}^{EckM}_{\pi N}$ is used.  In particular, the parameters obtained for
ordinary muon capture are a subset of those required for our calculation of
radiative muon capture.

The external nucleon fields in our calculation are renormalized by defining a
wave function renormalization factor, $Z_N$, which is the residue of the full
heavy baryon nucleon propagator at the pole.  Even though $Z_N$ is not
measurable, it affects measurable quantities, and one must use the value
consistent with the Lagrangian and the rest of the calculation. Our result
differs from that of previous published work using different Lagrangians, but
is the one appropriate for the new E\&M Lagrangian, as can be checked for
example by noting that our form is the one within our formalism that is
necessary to ensure that the vector coupling is not renormalized.

There is an additional normalization factor relating the normalization of
relativistic and heavy baryon wave functions, which we put in explicitly. An
alternative approach recently suggested \cite{ecker97} absorbs this momentum
dependent factor into the definition of $Z_N$. This gives a different
expression for $Z_N$ than we obtain, but would lead to exactly the same
physical results.

\section{NON RADIATIVE MUON CAPTURE}
The general amplitude for muon capture can be parameterized in terms of the
form factors $G_V(q^2)$, $G_M(q^2)$, $G_A(q^2)$ and $G_P(q^2)$ as follows,
\begin{displaymath} 
   {\cal M} = \frac{-iG_\beta}{\sqrt{2}}l_\alpha
            \overline{u}({\bf p}_n)\left[G_V(q^2)\gamma^\alpha
            +\frac{iG_M(q^2)}{2m_N}\sigma^{\alpha\beta}q_\beta
            -G_A(q^2)\gamma^\alpha\gamma_5
            -\frac{G_P(q^2)}{m_\mu}q^\alpha\gamma_5\right]u({\bf p}_p)~,
\end{displaymath}
where  $l_\alpha = \overline{u}({\bf p}_\nu)\gamma_\alpha
(1-\gamma_5)u({\bf p}_\mu)$ is the leptonic current, $m_N$ is 
the physical nucleon mass, and $G_\beta$ is the Fermi 
constant applicable to $\beta$-decay.

In HBChPT, the muon capture amplitude in terms of heavy baryon
spinors is
\begin{displaymath}
   {\cal M} = \frac{g_W}{2\sqrt{2}m_W^2}l_\alpha\overline{n}_v({\bf p}_n)
              \left(  \Gamma^{(r)\,\alpha}_{pWn}(q)
            + \Gamma^{(r)}_{p{\pi}n}(q)\left[\frac{i}{q^2-m_\pi^2}\right]
              \Gamma^{(r)\,\alpha}_{W\pi}(q)\right) n_v({\bf p}_p)~,
\end{displaymath}
with $m_\pi$ the physical pion mass, and $m_W, g_W$ the mass and weak coupling 
constant of the W boson.  

The functions $ \Gamma^{(r)\,\alpha}_{pWn}(q), \Gamma^{(r)}_{p{\pi}n}(q)$ and
$\Gamma^{(r)\,\alpha}_{W\pi}(q)$ are the fully renormalized vertex
functions. To evaluate them we start with the Lagrangian, ${\cal L}^{EckM}_{\pi
N} = \widehat{\cal L}^{(1)}_{\pi N} + \widehat{\cal L}^{(2)}_{\pi N} +
\widehat{\cal L}^{(3)}_{\pi N}$, where the ${\cal O}(p^3)$ part is the new part
of the Lagrangian as derived by E\&M. Using this Lagrangian the various
contributing diagrams are calculated and completely renormalized, so that they
can be expressed in terms of physical quantities. One then uses the relation
between the heavy baryon spinors $n({\bf p})$ and the Dirac spinors $u({\bf
p})$, i.~e.,
\begin{displaymath}
n_v({\bf p}) = \sqrt{\frac{2m_N}{m_N+v \cdot p_p}}
\frac{(1+v\!\!\!/)}{2}u({\bf p}) =\left[1-\frac{k\!\!/_p}{2m_N}+\frac{(m_N-
m_{0N})}{2m_N}+\frac{k_p^2}{8m_N^2}+ {\cal O}(\frac{1}{m_N^3})\right] 
u({\bf p}), 
\end{displaymath}
where $p=m_{0N}v + k_p$, to express the amplitude in the
original form and extract the form factors.
The result is

\pagebreak
  
\begin{eqnarray*}
   G_V(q^2) &=& 1 - \left(a_6-\frac{1}{8}\right)\frac{q^2}{m_N^2}
         - \frac{q^2}{18(4{\pi}F)^2}(1+17g_A^2) \\
     &&   - \frac{2q^2}{(4{\pi}F)^2}\left[
           b_7^r(\mu)+\frac{1}{12}(1+5g_A^2){\rm ln}\left(
           \frac{m_\pi^2}{\mu^2}\right)\right]  \\
     &&  + \frac{2}{(4{\pi}F)^2}\left[\frac{m_\pi^2}{3}(1+2
           g_A^2)-\frac{q^2}{12}(1+5g_A^2)
           \right]
           \int_0^1{\rm d}x\,{\rm ln}\left(1-x(1-x)\frac{q^2}{m_\pi^2}\right)
           \label{gV} \\
   G_M(q^2) &=& 4a_6 - 1 - \frac{4{\pi}g_A^2m_\pi{m}_N}{(4{\pi}F)^2}
     \int_0^1{\rm d}x\sqrt{1-x(1-x)\frac{q^2}{m_\pi^2}} \label{gM},\nonumber \\
   G_A(q^2) &=& g_A + \frac{4a_3g_Am_\pi^2}{m_N^2}
                - \frac{g^3_Am_\pi^2}{(4{\pi}F)^2} \nonumber \\
            &&  + \frac{4m_\pi^2}{(4{\pi}F)^2}\left[b_{17}^r(\mu)-\frac{g_A}{4}
                  (1+2g_A^2)\,{\rm ln}\left(\frac{m_\pi^2}{\mu^2}\right)\right]
                - \frac{b_{23}q^2}{(4{\pi}F)^2}, \label{gA} \nonumber \\
   G_P(q^2) &=& \frac{2m_{\mu}m_N}{(m_\pi^2-q^2)}\left[G_A(q^2)
           - \frac{m_\pi^2}{(4{\pi}F)^2}(2b_{19}-b_{23})\right] 
             \label{gP} \nonumber,
\end{eqnarray*}
where $F=$92.4 MeV is the pion decay constant.

The parameters appearing in these expressions can be evaluated by comparison
with other known experimental quantities. $G_A(0)$ can be obtained from data on
neutron decay, thus giving the parameter $g_A$ to the order needed. $G_M(0)$ is
related to the nucleon magnetic moments, which leads to the standard value of
$a_6$, one of the parameters of the ${\cal O}(p^2)$ Lagrangian.  The $q^2$
dependence of the nucleon electromagnetic form factors gives $G_V(q^2)$ and
thus the value of $b_7^r(m_N)$. Similarly the $q^2$ dependence of the axial
form factor as measured in antineutrino-nucleon scattering (or pion
electroproduction) gives an estimate of $b_{23}$. Finally $b_{19}$ is related
to the so-called Goldberger-Treiman discrepancy, and can be evaluated using the
pion nucleon coupling constant as input.

We thus find for the three constants of the E\&M Lagrangian obtainable from
this process $ b_7^r(m_N) = -0.53 \pm 0.02$, $ b_{23} = -3.1 \pm 0.3$, and $
b_{19} = -0.7 \pm 0.4$~.  Recently the constant $b_{19}$ has also been
determined by Moj\v{z}i\v{s} \cite{moj} in the context of $\pi$N
scattering. His reported value of $ b_{19} = -1.0 \pm 0.4$, corresponding to
$g_{\pi N}=13.0 \pm 0.1$, is consistent with our value of $b_{19}$ within
errors, with the difference being due almost entirely to his choice of a
different value of $F_\pi$.

We can now evaluate $ G_P$ using these values and we obtain $G_P(-0.88m_\mu^2)
= 8.21 \pm 0.09$ which is in good agreement with the best value from
non-radiative muon capture,\cite{bardin} $G_P(-0.88m_\mu^2) = 8.7 \pm 1.9$~.

Thus we have obtained the form factors of muon capture by a proton in the
framework of the recently derived E\&M ${\cal O}(p^3)$ heavy baryon chiral
Lagrangian, and used experimental data to extract numerical values for some of
the Lagrangian's parameters, which will be needed for the calculation of
radiative muon capture. Further details are given in \cite{full}.

\section{RADIATIVE MUON CAPTURE}
The radiative muon capture process, $\mu + p \rightarrow n + \nu + \gamma$ is
particularly interesting because it is especially sensitive to the value of the
induced pseudoscalar form factor, $G_P(q^2)$. A recent TRIUMF experiment
\cite{jonkmans} measured the rate for the radiative process and found a value
of $G_P$ approximately 1.5 times the value expected from the Goldberger-Treiman
relation.  While consistent with radiative capture measurements in nuclei, this
value is at variance with results from the nonradiative capture both in nuclei
and on the proton, which agree with the Goldberger-Treiman relation.

The standard theoretical calculation for radiative capture \cite{hwf} involves
four tree level Feynman graphs, consisting of radiation from the external
particles and from the intermediate pion generating $G_P$, together with a
gauge term generated by minimal substitution. Corrections due to $\Delta$
intermediate states have been calculated \cite{beder} but are small.

In view of the discrepancy with experiment it is of interest to apply the same
heavy baryon chiral perturbation theory approach used for ordinary capture to
the radiative capture process. This allows a microscopic calculation of the
gauge term and one would expect a number of new contributions for which there
is no counterpart in the standard Feynman graph approach.

We have approached this process using the same techniques used for ordinary
muon capture. In particular we use the same ${\cal O}(p^3)$ Lagrangian of Ecker
and Moj\v{z}i\v{s} \cite{EM} and our same expression for the wave function
renormalization factor \cite{full}. The strategy is to evaluate individually
the complete renormalized irreducible vertex functions for the separate
interactions, i. e. for the weak-NN, $\pi$NN, $\gamma$NN, $\gamma \pi \pi$ and
weak-$\pi$ vertices. We then put the pieces together to get the radiative muon
capture amplitude.

We expect that the results corresponding to the first four graphs of the
standard approach will be similar to the standard calculation. These graphs,
corresponding to radiation from external legs and from an intermediate pion,
involve situations where the electromagnetic and weak vertices are
separated. Thus any approach which uses the non radiative muon capture and
electromagnetic vertices to fix the parameters of the interaction, as we have
done, should pretty much reproduce the standard calculation of these
graphs. There may be possible off shell effects and the main terms and their
leading relativistic corrections may arise from various pieces of the ChPT
Lagrangian, but the basic physics is the same as in the standard approach.

This is not the case for the gauge term however, which is included in the
standard calculation only by way of a minimal substitution. In contrast, the
ChPT approach makes an explicit prediction for contributions to this diagram
and there are many such contributions. Some contain loops, with both weak and
electromagnetic vertices attaching to the loop, and some arise from contact
terms from the ${\cal O}(p^3)$ Lagrangian. Most of these contributions do not
appear in the standard minimal coupling diagram and, except that they are
${\cal O}(p^3)$ terms, one does not know {\em a priori} how large they will be.

We have explicitly evaluated one such contribution arising from the
Wess-Zumino-Witten part of the Lagrangian. This contribution leads to a diagram
with intermediate pion coupling at a point to weak and electromagnetic
currents. It contains the pion propagator and so contributes to $G_P$. An
analogous contribution was important for the non soft photon corrections to the
spin dependent part of the virtual Compton scattering amplitude
\cite{hemmert}. Like radiative muon capture, that process involves the coupling
to two external currents. Furthermore this term is gauge invariant by itself,
and so can be evaluated individually. Unfortunately it turns out to be
negligibly small, apparently because for radiative muon capture, unlike Compton
scattering, there are leading contributions from lower order parts of the
Lagrangian.

There are however many other diagrams involving both loops and contact terms
which may contribute. These can be combined into the renormalized, irreducible,
weak-$\gamma$NN and $\pi\gamma$NN vertices which are now being calculated. When
combined appropriately with the other vertices these will allow an explicit
comparison with the radiative muon capture data. \cite{note}


\begin{thebibliography}{99}

\bibitem{bkm} V. Bernard, N. Kaiser and U.-G. Mei{\ss}ner, Int. J. Mod. Phys.
                E {\bf 4}, 193 (1995). 
\bibitem{Ecker95} G. Ecker, Prog. Part. Nucl. Phys. {\bf 35}, 1 (1995).

\bibitem{Pich} A. Pich, Reports on Progress in Physics {\bf 58}, 563 (1995), 
               hep-ph/9502366.
\bibitem{EM}  G. Ecker and M. Moj\v{z}i\v{s}, Phys. Lett. B {\bf 365}, 312
               (1996).
\bibitem{ecker97} G. Ecker and M. Moj\v{z}i\v{s}, hep-ph/9705216.
\bibitem{moj} M.  Moj\v{z}i\v{s}, hep-ph/9704415.
\bibitem{bardin} G. Bardin {\it et. al.}, Phys. Lett. B {\bf 104}, 320 (1981).
\bibitem{full} H. W. Fearing, R. Lewis, N. Mobed, and S. Scherer,
Phys. Rev. D {\bf 56}, 1783 (1997), hep-ph/9702394.

\bibitem{jonkmans} G. Jonkmans {\it et. al.\/}, Phys Rev. Lett. {\bf 77}, 4512 
                 (1996), nucl-ex/9608005. 
\bibitem{hwf} H. W. Fearing, Phys. Rev. C {\bf 21}, 1951 (1980). 
\bibitem{beder} D. S. Beder and H. W. Fearing, Phys. Rev. D {\bf 35}, 
2130 (1987); {\bf 39}, 3493 (1989). 
\bibitem{hemmert} T. R. Hemmert, B. R. Holstein, G. Kn\"ochlein, 
and S. Scherer, Phys. Rev. Lett {\bf 79}, 22 (1997).
\bibitem{note} After this contribution  was originally submitted there 
appeared two related preprints dealing with radiative muon capture in ChPT. 
One \cite{MMK} evaluated only simple tree-level diagrams, while the 
other \cite{min} included single loops.
\bibitem{MMK} T. Meissner, F. Myhrer and K. Kubodera, nucl-th/9707019.
\bibitem{min} S. Ando and D.-P. Min, hep-ph/9707504.

\end{thebibliography}
\end{document}